%% file: letter.arxiv.tex
\documentclass[12pt]{article}
\pdfoutput=1
\usepackage{scicite}
\usepackage{times}
\usepackage[latin1]{inputenc}
\usepackage{graphicx}
\newenvironment{sciabstract}{%
\begin{quote} \bf}
{\end{quote}}

\newcounter{lastnote}
\newenvironment{scilastnote}{%
\setcounter{lastnote}{\value{enumiv}}%
\addtocounter{lastnote}{+1}%
\begin{list}%
{\arabic{lastnote}.}
{\setlength{\leftmargin}{.22in}}
{\setlength{\labelsep}{.5em}}}
{\end{list}}

\title{Correlation of the highest energy cosmic rays with nearby extragalactic objects}
\author
{The Pierre Auger Collaboration\footnote{The full list of authors and their
affiliations appears at the end of this paper.}
\\
\normalsize{
Observatorio Pierre Auger, Avenida San Mart{\'\i}n Norte 304,}\\
\normalsize{ (5613) Malarg\"ue, Mendoza, Argentina}\\
\\
}

\date{}

\begin{document} 
\baselineskip24pt
\maketitle

\begin{sciabstract}
Using data collected at the Pierre Auger Observatory during the past 3.7 
years, we demonstrated a correlation between the 
arrival directions of cosmic rays with energy above  $\mathbf{\sim 6\times 
10^{19}}$~electron volts  and the positions of
active galactic nuclei (AGN) lying within $\mathbf{\sim 75}$~megaparsecs. 
We rejected the hypothesis of an isotropic distribution of these cosmic rays with
at least a 99\% confidence level from a prescribed a priori test.
The correlation we observed is compatible with the hypothesis
that the highest energy particles originate from nearby extragalactic sources
whose flux has not been substantially reduced by interaction with the 
cosmic background radiation. AGN or objects having a 
similar spatial distribution are possible sources. 
\end{sciabstract}

Cosmic rays are energetic particles and nuclei from space that strike the Earth's atmosphere. Their energies vary from a few $10^8$~eV to beyond $10^{20}$~eV. The flux of cosmic rays
at Earth decreases very rapidly with energy, from a few particles per square centimeter per second in the low-energy region to less than one particle per square kilometer per century above $10^{20}$~eV. The identification of the sources of ultrahigh-energy cosmic rays (UHECR) with energies $\sim 10^{20}$~eV has been a great challenge since they were first observed in 1962~\cite{linsley}.   Because cosmic rays at these energies are not expected to be confined by magnetic fields in the disk of our galaxy, and indeed no significant excess from the direction of the Milky Way has been observed, it is likely that they originate outside the Galaxy. Until now there has been no experimental confirmation of this hypothesis.

\begin{figure}[t]\centerline{\includegraphics[width=0.8\textwidth]{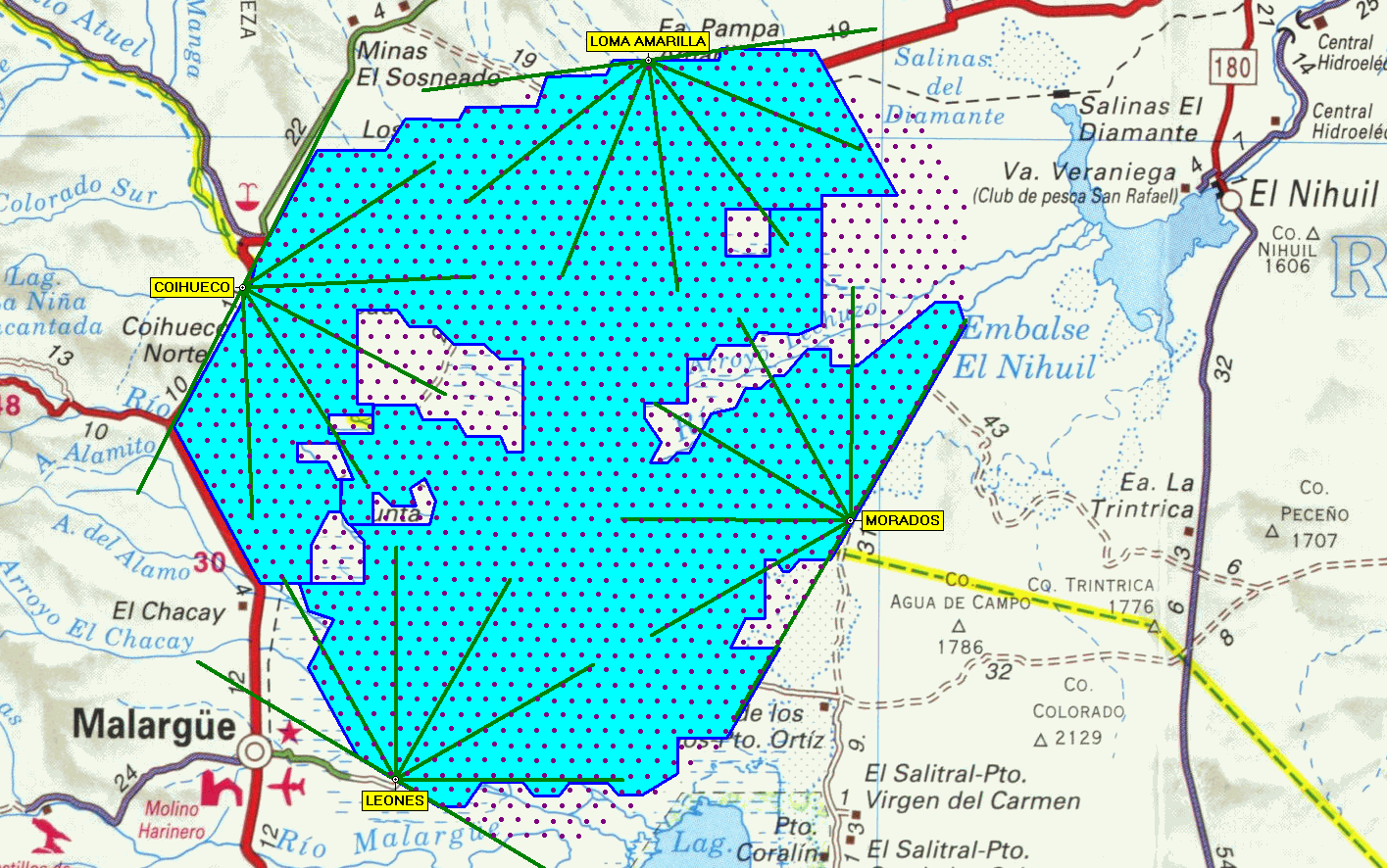}}\caption{Layout of the Pierre Auger Southern Observatory.  The dots represent the position of each of the 1600 SD stations. The 1430 SD stations deployed and activated as of 30 September 2007 lie in the area shaded blue. The 4 FD sites are labeled in yellow, with green lines indicating the field of view of the six telescopes at each site.  To give the scale of the Observatory, the lengths of the green line correspond to 20~km.}\label{array}\end{figure}

Because of their very low flux,  UHECR can only be detected through their interaction with the Earth's atmosphere, producing a cascade of billions of particles that excite nitrogen molecules in the air along their path and spread over 
a large area when they reach the ground.
The Pierre Auger Southern Observatory \cite{nim}, now nearing
completion in  Argentina, was designed to  simultaneously observe the shower particles at 
ground level and the associated fluorescence light generated in the atmosphere. 
A large array of 1600 surface detectors (SDs), laid out as an equilateral triangular grid with 1500-m spacing,  covers an area of 3000~km$^2$ and  detects the particles at ground level by means of the Cherenkov radiation they produce in water. At each of four sites on the periphery of the instrumented area, six inward-facing optical telescopes observe the sky on clear moonless nights.  These devices measure the atmospheric fluorescence light produced as an extensive air shower passes through the field of view.  The two techniques -- the SDs and the fluorescence detectors (FDs) -- are complementary, and also provide cross-checks and redundancy in the measurement of air shower parameters. The SD measures the two-dimensional lateral structure of the shower at ground level, whereas the FD records  the longitudinal profile of the shower during its  development through the atmosphere. In Figure~\ref{array}, we present the layout of the Observatory as of 30 September 2007.

The Pierre Auger Southern Observatory  has been taking data stably  since 
January 2004.  
The large 
exposure of its 
ground array, combined 
with accurate energy and arrival direction measurements,
calibrated and verified from the hybrid operation  with the fluorescence detectors, 
provides an opportunity to explore
the spatial correlation between cosmic rays  and their sources in the sky. 

If cosmic rays with the highest energies are predominantly protons or nuclei, only sources 
closer than about 200~Mpc from Earth can contribute appreciably to the observed flux above
60~EeV (1~EeV = $10^{18}$~eV). Protons or nuclei with energies above 60~EeV interact with the 
cosmic microwave 
background~\cite{GZK1,GZK2,GZK3},
leading to a strong attenuation of their flux from distant sources. This attenuation is known as the Greisen, Zatsepin and Kuzmin (GZK) effect, from the names of the three physicists that predicted it. 
If the sources of the most energetic cosmic rays are relatively nearby and are not uniformly 
distributed, then an anisotropic arrival distribution is expected, provided the particles 
have a sufficiently small charge and a sufficiently high energy for their directions to be 
minimally perturbed by intervening magnetic fields.

Anisotropy of the cosmic rays with the highest energies could
manifest as clustering of events from individual point sources or
through the correlation of arrival directions with a collection of
astronomical objects. The Akeno Giant Air Shower Array (AGASA) Collaboration claimed some 
excess of clustering at small angular scales compared to 
isotropic expectations~\cite{agasa}, but this was not supported
by data recorded by the HiRes experiment~\cite{Hires}.
Analyses of data recorded by several air-shower experiments
revealed a general correlation with the direction of the
supergalactic plane \cite{sgp1,sgp2}, where several nearby galaxies
cluster, but with limited statistical significance.

AGN  have long been considered sites where energetic-particle production
might take place
and where protons and heavier nuclei could be 
accelerated up to the highest energies yet measured  
\cite{ginzburg,hillas}. Here, we report the 
observation of a correlation between the arrival 
directions of the cosmic rays with highest energies measured by the 
Pierre Auger Observatory  and the positions of nearby AGN
from the $12^{th}$  edition of the catalog of quasars and active nuclei
by V\'eron-Cetty and V\'eron  (V-C catalog)~\cite{VC06}. 

\paragraph*{Data set and method}
 The data set analyzed here consists of the cosmic-ray events  recorded 
 by the surface array of the Observatory  from 1 January 2004 to
 31 August 2007. It contains 81 events with reconstructed energies above 40 EeV 
and zenith angles smaller than 
$60^\circ$. The integrated exposure is  $9.0\times 
10^{3}$~km$^2$~sr~year.

We only use recorded events if they meet strict criteria with regard to the quality of the reconstruction of their energy and direction. The selection of those events is done via a quality trigger~\cite{cuts} which is only a function of the topology of the footprint of the event on the ground. This trigger requires that the detector with the highest signal must be surrounded by five active nearest neighbors, and that the reconstructed shower core be inside an active equilateral triangle of detectors. This represents an efficient quality cut while guaranteeing that no crucial information 
is missed for the shower reconstruction. 

The arrival direction of a cosmic ray is a crucial ingredient in our study. 
The event direction is determined by a fit of the arrival times of the shower front at the SD. 
The precision achieved in the arrival direction depends on the clock
resolution of each detector and on the fluctuations in the time of arrival of the first
particle~\cite{timevariance}. The angular resolution is defined as the angular
aperture around an arrival direction of cosmic rays within which 68\% of
the showers are reconstructed. This resolution has been verified experimentally
with events for which two independent geometrical reconstructions can be performed. 
The first test uses hybrid events, which are measured simultaneously by  the SD and the FD; the second one uses events falling in a special region of our array where two surface stations are laid in pairs 11~m
apart at each position. 
Events that triggered at least six surface stations have energies above 10~EeV and an  angular resolution
better than $1^\circ$~\cite{carla,ave}.

The energy of each event is determined in a two-step procedure. The shower size $S$, at a 
reference distance and zenith angle, is calculated from the signal detected in each surface station and then converted to energy with a linear calibration curve based on the fluorescence telescope measurements~\cite{roth}. The uncertainty resulting from the adjustment of the shower size, the conversion to a reference angle, the fluctuation from shower to shower, and the calibration curve amounts  to about 18\%.  
The absolute energy scale is given by the fluorescence measurements and has a systematic uncertainty of 22\%~\cite{bruce}.  
The largest systematic uncertainty arises primarily from an incomplete knowledge of the yield of photons from the
fluorescence of atmospheric nitrogen (14\%), the telescope calibration (9.5\%) and the reconstruction procedure (10\%).  
Additional uncertainty in the energy scale for the set of high-energy
events used in the present analysis is due to the relatively low
 statistics available for calibration in this energy range.

Events with energy above 3 EeV are recorded with nearly 100\% efficiency over the area covered by the 
surface array. The nonuniformity of the exposure in right ascension is below 1\%,  negligible in 
the context of the present analysis. 
The dependence of the exposure
on declination is calculated from the
latitude of the detector and the full acceptance for showers up to 
$60^{\circ}$
zenith angle.

A key element of our study is the probability $P$ for 
a set of $N$ events from an isotropic flux to contain $k$ or more events at a
maximum angular distance $\psi$ from any member of a collection 
of candidate point sources. ${\mathbf P}$ is given by the cumulative binomial distribution 
$\sum_{j=k}^N C^N_j p^j(1-p)^{N-j}$, where  the parameter $p$
is  the fraction of the sky (weighted by the exposure) defined by the regions 
at angular separation less than  $\psi$ from the selected sources.  

We analyze the degree of correlation of our data  with the directions of AGN 
referenced in the V-C catalog \cite{VC06}. 
This catalog does not contain all existing AGN 
and is not an unbiased statistical sample of them.  This is not an obstacle to demonstrating
the existence of anisotropies but may affect our ability to identify 
the cosmic-ray sources unambiguously.
The catalog contains 694 active galaxies with  redshifts $z\le 0.024$, corresponding to distances
$D$ smaller than 100 Mpc~\cite{note:1}. At larger distances, and around the Galactic plane, 
the catalog is increasingly incomplete.

\paragraph*{Exploration and confirmation}
Using data acquired between 1 January 2004 and  26 May 2006, we scanned for the minimum of 
$P$
in the three-dimensional parameter space defined by maximum angular separations 
$\psi$,
maximum redshifts $z_{max}$,
and energy thresholds $E_{th}$.
The lower limit for the scan in $\psi$ 
corresponds to the angular resolution of the surface array. 
Our scan in energy threshold and maximum distance was motivated by the assumption
that cosmic rays with the highest energies are the ones that are least
deflected by intervening magnetic fields and that have the smallest
probability of arrival from very distant sources due to the GZK effect
\cite{GZK1,GZK2}. 
 
We found a minimum of $P$
for the parameters $\psi=3.1^\circ$, 
$z_{max}=0.018$ ($D_{max}= 75$~Mpc), and $E_{th} = 56$~EeV. 
For these values,
12 events among 15  correlate with the selected AGN, 
whereas only 3.2 were expected by chance if the flux were isotropic. 
This observation
motivated the definition of a test to validate the result 
with an independent data set, with parameters specified a priori, as is required by the Auger 
source and anisotropy 
search methodology~\cite{clay,revenu}. 

The Auger search protocol was designed as a sequence of tests to be applied after the
observation of each new event with energy above 56~EeV. The 
total probability of incorrectly rejecting the isotropy hypothesis along the sequence was set to 
a maximum of 1\%. 
The parameters  for the prescribed test were chosen as those, given above,
that led to the minimum of $P$ in the exploratory scan. 
The probability of a chance correlation at the chosen angular
scale of a single cosmic ray
with the selected astronomical objects 
is $p=0.21$ if the flux were isotropic. 
The test was applied 
to data collected between 27 May 2006 and 31 August 2007, with exactly the same reconstruction algorithms, 
energy calibration, and quality cuts for event selection as in the exploratory scan.
In these independent data, 
there are 13 events with energy above 56~EeV, of which 8 have arrival directions 
closer than 3.1$^\circ$ from the positions of AGN less than 75 Mpc away, 
with 2.7 expected on average.
The probability that this configuration would occur
by chance if the flux were isotropic is $1.7\times 10^{-3}$.
Following our search protocol  and based on the independent data set alone, we reject 
the hypothesis of isotropy in the distribution of  the arrival directions of cosmic rays 
with the highest energies with at least a 99\% confidence level.

\paragraph*{Results}
Having determined that an anisotropy exists, based on the a priori prescription, we  
rescanned
the full data set from 1 January 2004 to 31 August 2007 using the method described above
to substantiate the observed correlation.
We used steps of $0.1^\circ$ in $\psi$, in the range
$1^\circ\le \psi \le 8^\circ$, 
and $0.001$ in $z_{max}$, in the range $0\le z_{max}\le 0.024$. 
We also used a newer version of our reconstruction and calibration algorithm that
gives slightly different reconstructed directions and energies.
These small differences, well within our reconstruction uncertainty, modify the final 
event selection, but this has minor consequences on the value of the parameters $\psi$, 
$z_{max}$, and $E_{th}$ that maximize the correlation signal.  
We start the scan with the event of highest energy and add events 
one by one in order of decreasing energy, down to $E_{th} = 40$~EeV.

Strong correlation signals occur for energy thresholds around
60 EeV and several combinations of the other parameters in the range
$\psi \le 6^\circ$, and $z_{max}\le 0.024$ ($D_{max} <$~100~Mpc).
The absolute minimum value of $P$ occurs for the 27 events with the highest energies (above 57~EeV in the new analysis).
We generated simulated sets of
directions, drawn from an isotropic distribution in proportion to
the relative exposure of the observatory. Performing an identical scan
on those simulated samples to that applied to the real data, we obtain
smaller or equal values of $P$ in 
$\sim 10^{-5}$ of the simulated 
direction sets.

\begin{figure}[t]
\centerline{\includegraphics[width=1.05\textwidth]{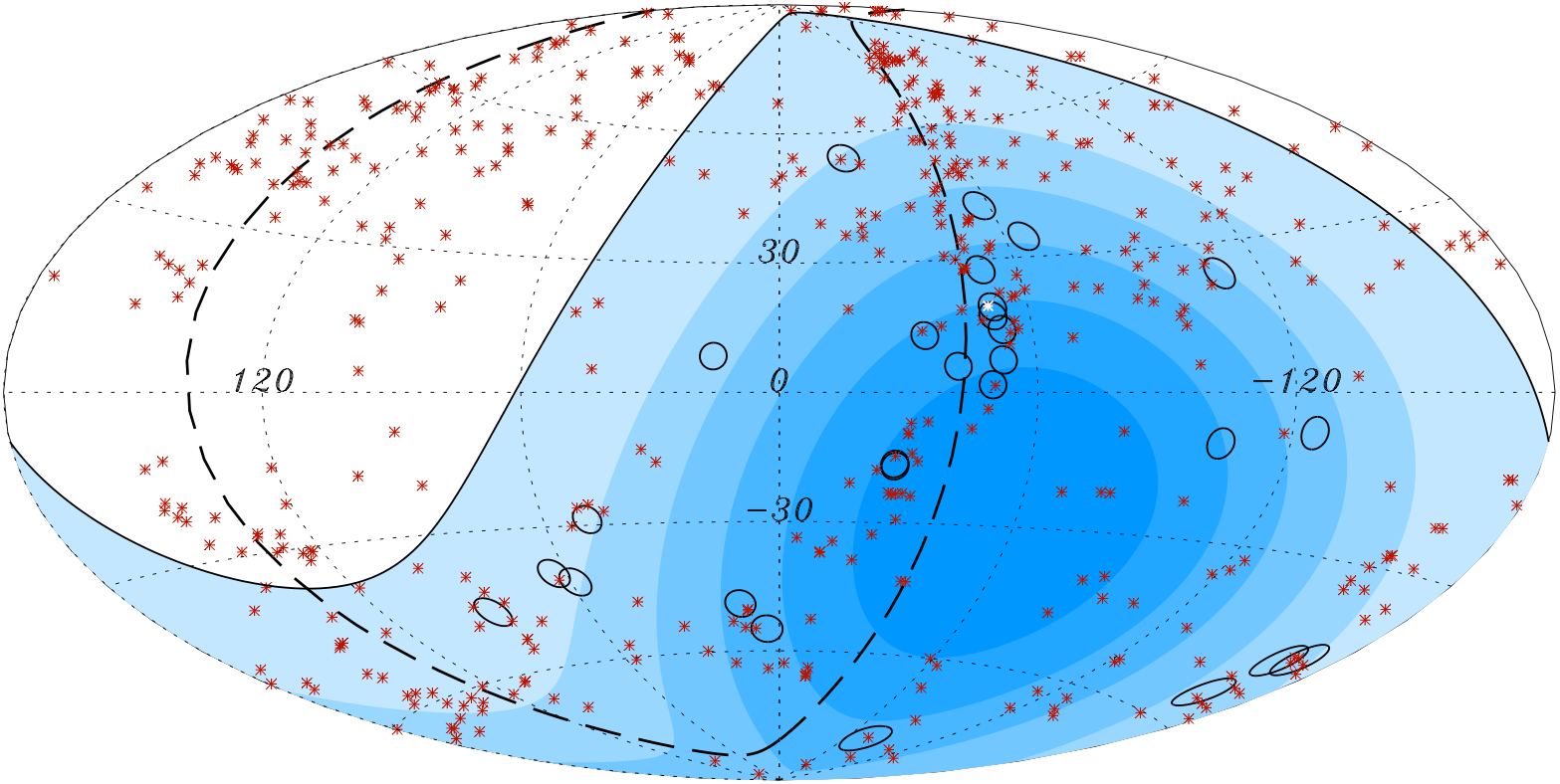}}
\caption{
Aitoff projection of the celestial sphere in galactic coordinates
with circles of radius $3.1^\circ$
centered at the arrival directions of the 27 cosmic rays with highest energy
 detected by the
Pierre Auger Observatory. The positions of the
472 AGN (318 in the field of view of the Observatory) with redshift $z\le 0.018$ ($D 
< 75$ Mpc)  from the $12^{th}$ edition of the catalog of
quasars and active nuclei \cite{VC06}
are indicated by red asterisks.
The solid line represents the border of the field of view (zenith angles smaller than $60^\circ$).
Darker color indicates larger relative exposure. Each colored band has equal integrated
exposure. The
dashed line is  the supergalactic plane. Centaurus A,
one of our closest AGN, is marked in white.
} \label{skymap}
\end{figure}

We present (Figure~\ref{skymap}) a sky map in Galactic coordinates of our 
27 highest-energy events ($E > 57$~EeV), as determined by our most recent
 version of the reconstruction code.  
The anisotropy is clearly visible. 
We note the proximity of several events close to the supergalactic
plane, and also that two events arrive within 3$^\circ$ of
Centaurus A, one of the closest AGN, marked in white on the figure.

\paragraph*{Discussion}
With the statistics of our present data set, the observed correlation 
is significant  for 
maximum distances to AGN  of up to 100~Mpc, for maximum angular separations 
of  up to $6^\circ$, and for energy thresholds 
around 60~EeV. Those numbers are to be taken as indicative because
the minimization of $\mathbf P$ is not totally exempt from biases.  Accidental correlation 
with foreground AGN different from the actual sources may induce bias towards 
smaller maximum source distances while accidental correlation with distant background 
ones may reduce the optimal maximum angular separation by a 
few degrees. 

Under the simplifying assumptions of a uniform distribution of sources 
with equal intrinsic luminosity and continuous energy loss in the 
cosmic microwave background due to the GZK effect~\cite{GZK1,GZK2}, 
90\% of the protons arriving at Earth with energy exceeding  60~EeV  originate 
from sources closer than 200~Mpc. 
This (somewhat arbitrarily defined) ``GZK horizon'' decreases rapidly
with increasing energy and drops to 90 Mpc for energies exceeding 
80 EeV.
The relation between the horizon distance and the value of
$D_{max}$ that minimizes P is not a simple one, given the possible
biases in the method, which has nonuniform sensitivity over
the range of parameters scanned. Increasing catalog incompleteness
also prevents confidently scanning over sources at distances much larger than
100 Mpc. Moreover, the local density
and luminosities of sources could have significant departures from the
uniformity assumed in the GZK horizon scale for a given energy
threshold. Taking into consideration these caveats, in addition to the
uncertainty in the reconstructed energies, the range
of $D_{max}$ and $E_{th}$ over which we observe a significant correlation
 is compatible with the frequently made assumption that the highest energy
cosmic rays are protons experiencing predicted GZK energy losses.  We note
that the correlation increases abruptly at
the energy threshold of 57~EeV, which coincides with the point on
the energy spectrum recently reported from the observatory at which
the flux is reduced by $\sim 50$\% with respect to a power law
extrapolation of lower-energy observations \cite{roth}.
 
If the regular component of the galactic magnetic field is coherent over
scales of  1 kpc with a strength of a few $\mu$G, 
as indicated by data from studies of pulsars~\cite{MAGP},
the observed correlation over an angular scale of only a few degrees for 
$E\sim$~60~EeV is indicative that 
most of the primaries are not heavy nuclei.

These features are compatible with  the interpretation that the 
correlation we observe
is evidence for  the GZK effect 
and the hypothesis that the highest-energy cosmic  rays reaching Earth are mostly protons from 
nearby sources. 

The catalog of AGN that we use is increasingly incomplete near the Galactic plane,
where
extinction from dust in the Milky Way reduces the sensitivity of
observations.
Deflections from the Galactic magnetic field are also expected to be 
significantly larger than average for cosmic rays that arrive at 
equatorial
Galactic latitudes, because they traverse a longer distance across any
regular Galactic magnetic component. These effects
are likely to have some impact upon the estimate of the strength of 
the correlation. Six out of the eight events that do not correlate 
with AGN positions within our prescribed parameters and reconstruction code
 lie less than 12$^\circ$  away from the Galactic plane. 

Despite its strength, the correlation that we observe with nearby AGN from the 
V-C catalog
cannot be used alone as a proof that AGN are the sources. 
Other sources, 
as long as their distribution within the GZK horizon is sufficiently similar to that of the AGN, could lead to a significant 
correlation between the arrival directions of cosmic rays and the AGN positions.  
Such correlations are under investigation in particular for the Infra-Red Astronomical
Satellite (IRAS) galaxies.
The autocorrelation signal of the highest-energy events is also being investigated. It
shows departures from isotropic expectations at angular scales between 5$^\circ$ and 20$^\circ$~\cite{silvia}
and serves as an additional tool to identify the spatial distribution of the sources.

\paragraph*{Conclusion}

We have demonstrated the anisotropy of the arrival directions of the highest-energy cosmic rays and their extragalactic origin. Our observations are consistent with the  hypothesis that the rapid decrease of flux measured  by the Pierre Auger Observatory above 60~EeV is due to the  GZK effect and  that most of the  cosmic rays reaching Earth in that energy range are protons from nearby astrophysical sources, either AGN or other objects with a similar spatial distribution.

The number of high-energy cosmic-ray events recorded so far by the
Pierre Auger Observatory and analysed in this work corresponds to
1.2 years of operation of the complete southern array.
The data set that the observatory will gather in just a few more years should offer a better 
chance to unambiguously identify the sources.
The pattern of correlations of cosmic-ray events with their sources could also 
assist in determining the properties of the intervening magnetic-field structures and in particle physics 
explorations at the largest energies. Astronomy based on cosmic rays with the 
highest energies opens a new window 
on the nearby universe.

\renewcommand\refname{References and Notes}

\begin{scilastnote}
\item 
We are grateful to the following agencies and organizations for financial support:  
Gobierno De La Provincia de Mendoza, Comisión Nacional de Energía Atómica, Municipalidad de Malargüe, Fundación Antorchas, Argentina; the Australian Research Council; Conselho Nacional de Desenvolvimento Científico e Tecnológico (CNPq), Financiadora de Estudos e Projetos do Ministerio da Ciencia e Tecnologia (FINEP / MCT), Fundação de Amparo à Pesquisa do Estado de Rio de Janeiro (FAPERJ), Fundação de Amparo à Pesquisa do Estado de São Paulo (FAPESP), Brazil; Ministry of Education, Youth and Sports of the Czech Republic; 
Centre National de la Recherche Scientifique (CNRS), Conseil Régional Ile-de-France, Département  Physique Nucléaire et Corpusculaire (PNC-IN2P3/CNRS), Département Sciences de l'Univers (SDU-INSU/CNRS), France; Bundesministerium für Bildung und Forschung (BMBF), Deutsche Forschungsgemeinschaft (DFG), Finanzministerium Baden-Württemberg, Helmholtz-Gemeinschaft Deutscher Forschungszentren (HGF), Ministerium für Wissenschaft und Forschung, Nordrhein Westfalen, Ministerium für Wissenschaft, Forschung und Kunst, Baden-Württemberg, Germany; Istituto Nazionale di Fisica Nucleare (INFN), Ministero dell'Istruzione, dell'Università e della Ricerca (MIUR), Italy; Consejo Nacional de Ciencia y Tecnología (CONACYT), Mexico; Ministerie van Onderwijs, Cultuur en Wetenschap, Nederlandse Organisatie voor Wetenschappelijk Onderzoek (NWO), Stichting voor Fundamenteel Onderzoek der Materie (FOM), Netherlands; Ministry of Science and Higher Education, Poland; Fundação para a Ciência e a Tecnologia, Portugal; Ministry for Higher Education, Science, and Technology, Slovenian Research Agency, Slovenia; Comunidad de Madrid, Consejería de Educacíon de la Comunidad de Castilla La Mancha, FEDER funds, Ministerio de Educacíon y Ciencia, Xunta de Galicia, Spain; Science and Technology Facilities Council, United Kingdom; Department of Energy, National Science Foundation, The Grainger Foundation, USA;  América Latina Formación Académica - European Community / High Energy physics Latin-american European Network (ALFA-EC / HELEN), and UNESCO.
\end{scilastnote}

\clearpage

\clearpage
\include{author_list}
\end{document}

%% file: author_list.tex
\par\noindent
{\bf The Pierre Auger Collaboration} \\
J.~Abraham$^{6}$, 
P.~Abreu$^{59}$, 
M.~Aglietta$^{45}$, 
C.~Aguirre$^{8}$, 
D.~Allard$^{24}$, 
I.~Allekotte$^{1}$, 
J.~Allen$^{78}$, 
P.~Allison$^{80}$, 
C.~Alvarez$^{49}$, 
J.~Alvarez-Mu\~{n}iz$^{65}$
% ...

\par\noindent
$^{1}$ Centro At\'{o}mico Bariloche (CNEA); Instituto Balseiro 
(UNCuyo), R\'{\i}o Negro, Argentina \\
% ...
$^{6}$ Universidad Tecnol\'{o}gica Nacional, Regionales Mendoza y 
San Rafael, Mendoza, Argentina \\
% ...
$^{8}$ Universidad Catolica de Bolivia, La Paz, Bolivia \\
% ...
$^{24}$ Laboratoire AstroParticule et Cosmologie, Universit\'{e} 
Paris 7, IN2P3/CNRS, Paris, France \\
% ...
$^{45}$ Istituto di Fisica dello Spazio Interplanetario (INAF),
 Universit\`{a} di Torino and Sezione INFN, Torino, Italy \\
% ...
$^{49}$ Benem\'{e}rita Universidad Aut\'{o}noma de Puebla, Puebla, 
Mexico \\
% ...
$^{59}$ LIP and Instituto Superior T\'{e}cnico, Lisboa, Portugal \\
% ...
$^{65}$ Universidad de Santiago de Compostela, Spain \\
% ...
$^{78}$ New York University, New York, NY, USA \\
% ...
$^{80}$ Ohio State University, Columbus, OH, USA \\
% ...